\documentclass[12pt]{article} 
\pdfoutput=1
     \usepackage{amsmath}
  \usepackage{graphicx}
  \usepackage{color}
  \usepackage{subfigure}
  \usepackage{tikz}
  \newlength{\abstractwidth}
 \usepackage[numbers,sort&compress]{natbib}
 \usepackage{amsmath,amssymb}
\usepackage{graphicx}
\usepackage{caption}
\usepackage{lipsum}
\usepackage[utf8]{inputenc}
 \usepackage{mciteplus} 

\usepackage[colorlinks = true,
            linkcolor = blue,
            urlcolor  = blue,
            citecolor = blue,
            anchorcolor = blue]{hyperref}

\hypersetup{colorlinks=true,linkcolor=blue}

\usepackage[title]{appendix}
\usepackage{comment}
\usepackage{braket}
\usepackage{tensor}

\makeatletter
\def\blfootnote{\xdef\@thefnmark{}\@footnotetext}
\makeatother

\newcommand{\R}{\tensor{R}}

\newcommand{\F}{\tensor{F}}
\newcommand{\T}{\tensor{T}}
\newcommand{\tGamma}{\tensor{\Gamma}}

\begin{document}

\begin{titlepage}
  \bigskip

  \bigskip\bigskip

  \bigskip

\begin{center}
{\Large \bf{}}
 \bigskip
{\large \bf {Revisiting $R^4$ higher curvature corrections to black holes}} 
\bigskip
\bigskip
   \bigskip
\bigskip
\end{center}

  \begin{center}

 \bf {Yiming Chen}
  \bigskip \rm
\bigskip
 
 \rm

Jadwin Hall, Princeton University,  Princeton, NJ 08540, USA\\

  \end{center}

 \bigskip\bigskip

\begin{abstract}

We revisit the corrections to black holes due to the $R^4$ terms in the
action. We discuss corrections to the metric and possible scalar fields,
as well as corrections to thermodynamic quantities. We also comment on the large $D$ limit of the solutions.

 \end{abstract}
\bigskip \bigskip \bigskip
\blfootnote{ymchen.phys@gmail.com}

  \end{titlepage}



\section{Introduction}

In this note, we compute the corrections to black holes due to an $R^4$ higher derivative term in the action. The particular higher derivative term we consider arises as the leading higher curvature correction in the low energy effective theory of both the type II superstring theory and the M-theory. 

In \cite{Myers:1987qx} (see also \cite{Moura:2018psq}), the corrections to the Schwarzschild solution due to this term were computed. We will follow its method, and with the help of mathematica, we correct some precise numerical coefficients  in it.  
More concretely, we will be considering the Schwarzschild solutions under the effective action
\begin{equation}\label{action10d}
        I = \frac{1}{16\pi G_N} \int d^{10} x  \sqrt{-g} e^{-2\phi} \left( R +4 (\nabla \phi)^2 - \frac{1}{2} e^{2\phi} (\nabla \chi)^2    +  z (\phi, \chi)  Y(R)\right),
\end{equation}
where $\phi$ and $\chi$ are the dilaton and the axion, respectively. $Y( R)$ is an eight derivative tensor composed of the Riemann tensor
\begin{equation}
     Y(R) = 2 R_{hmnk} \R{_p^m^n_q }  R^{hwsp}\R{^q_w_s^k} +  R_{hkmn} \R{_p_q^m^n}  R^{hwsp}\R{^q_w_s^k},
\end{equation}
The correction term in (\ref{action10d}) appears at $\mathcal{O}(\alpha'^3 )$, which makes the computation more difficult than the bosonic case \cite{Callan:1988hs}, where the leading correction is at $\mathcal{O}(\alpha')$. In the action (\ref{action10d}), we have left out the supersymmetric completion as well as the gauge fields, which will not be turned on in the solutions we study. In general, the form of the eight derivative term can be modified by field redefinitions \cite{Gross:1986iv,Tseytlin:1986zz}. Here we choose to work in the scheme that the tensor contains only the Riemann tensor, but not the Ricci scalar or Ricci tensor. The effective action can be derived either from the sigma model beta functions \cite{Grisaru:1986vi,Freeman:1986zh,Park:1987jp}, or by inspecting the four-graviton scattering amplitude \cite{Gross:1986iv,Myers:1987qx}. In principle, the $Y(R)$ term should also involve derivative couplings to the dilaton, but they vanish at least to the linear order in $\phi$ \cite{Myers:1987qx}, so we can ignore them in our perturbative calculation as we'll see below.

The coupling $z(\phi,\chi)$ in front of the $Y(R)$ term equals to a constant $\frac{1}{16}\zeta(3) \alpha'^3$ from the tree level computation, while more generally it is a function of the dilaton $\phi$ and axion $\chi$ as it receives higher loop and nonperturbative corrections (see e.g. \cite{Chamseddine:1991qu,Green:1997tv,Green:1997di}). The explicit form of $z(\phi,\chi)$ depends on the details of the theory, but we will leave it general here. 
It is also worth noting that as a general low energy effective theory, the value of $z$ can be bounded by the S-matrix bootstrap method, see \cite{Guerrieri:2021ivu} for a recent progress. The effect of the $Y(R)$ term on the black hole solutions was also studied in some other contexts, for example black holes in AdS$_5\times S^5$ \cite{Gubser:1998nz,Pawelczyk:1998pb} and in M-theory in 11 dimensions \cite{Hyakutake:2013vwa}.

Our derivation in sec. \ref{sec:solution} will closely follow the notations and steps in \cite{Myers:1987qx}, so that the results can be easily combined by the reader. 
We also add some further discussions on the thermodynamics of the black hole in sec. \ref{sec:thermodynamics}.

This note is first motivated by a relation \cite{Emparan:2013xia,Chen:2021emg} between black holes in string theory in the large dimension limit and the two dimensional black hole \cite{Witten:1991yr,Dijkgraaf:1991ba}. As an application and also an indirect check of our computation, we comment in sec. \ref{sec:largeD} on the large dimension limit of the solution and its relation to the two dimensional black hole.

\section{Corrections to the black hole solution}\label{sec:solution}

We will be considering black holes in $D$ dimensions. If we were to consider it in the type II superstring theory, which is defined in 10 dimensions, we can take the rest of the $(10 - D)$ dimensions to be decoupled and satisfy its own set of equations. We can take the internal dimensions to be a flat torus for simplicity. In fact, for our discussion in sec. \ref{sec:largeD}, we will also be interested in the case with $D>10$, in which case we can add can add an extra dimension with a timelike linear dilaton, which gives negative central charge to make the theory consistent.\footnote{To have a standard GSO projection we need the total dimensions to be $8n + 2, \, n \in \mathbb{Z}$  \cite{Chamseddine:1991qu}.} Of course, in the case with an extra timelike direction, we should only consider the $D$ dimensional black hole as an Euclidean background in order to maintain unitarity.\footnote{We thank Arkady Tseytlin for questions related to this point.} 

The equations of motion following from the action are
    \begin{equation} \label{eom1}
          R +   4 \nabla^2 \phi   = - z Y + \frac{1}{2} \partial_\phi z Y
    \end{equation}
\begin{equation}\label{eom2}
    R_{ab} + 2 \nabla_a \nabla_b \phi = z T_{ab} + \frac{1}{4}g_{ab} \partial_\phi z  Y ,\quad T_{ab} \equiv -\frac{\delta Y(R)}{\delta g^{ab}}.
\end{equation}
    \begin{equation}\label{eom3}
        \nabla^2 \chi  = -  e^{-2\phi} \partial_\chi z Y,
    \end{equation}
where we've ignored terms involving $(\nabla \phi)^2 , (\nabla \chi)^2$, since we'll see they don't contribute to the leading order in perturbation. The variation $\frac{\delta Y(R)}{\delta g^{ab}}$ is given by
\begin{equation}\label{variationformula}
\begin{aligned}
   \frac{\delta Y(R)}{\delta g^{ab}} = &  2 R_{mhk(a} \R{_{b)}_w_p^m} \left( R^{kqsw} \R{^p_q_s^h} + R^{ksqp}\R{^h^w_q_s }\right) + 2 R_{kqs(a} R_{b)wmp} \left( R^{hsqp} \R{^k^w^m_h} - R^{phsq} \R{_h^w^m^k}\right) \\
    & - 2 \nabla_{(k} \nabla_{h)} \left( R_{apwb}R^{ksqp} \R{^h_s_q^w} \right) - 2 \nabla_{(w}\nabla_{m)} \left( R_{asqp} \R{_b^s^q_k} R^{wkpm}\right) \\
    & + 2 \nabla_{(k}\nabla_{s)} \left( \R{^h^w^s_{(a}} R_{b)mnw} \R{_h^m^n^k}\right) + 2 \nabla_{(k}\nabla_{s)} \left( \R{^s^h^p_{(a}} R_{b) mnp}  \R{^k^m^n_h}\right),
\end{aligned}  
\end{equation}
which we derive in appendix. \ref{app:variation}. This explicit expression can be found in \cite{Grisaru:1986vi}.\footnote{There is an overall sign difference from \cite{Grisaru:1986vi} due to different conventions. The convention we use is $\R{^d_c_a_b} u^c = [\nabla_a, \nabla_b] u^d$, $R_{ab} = \R{^c_a_c_b}$, and $T_{(ab)} = T_{ab} + T_{ba}$. } The case considered in \cite{Myers:1987qx} corresponds to setting $\partial_\phi z = \partial_\chi z= 0$.   

The Schwarzschild solutions at zeroth order are
\begin{equation}\label{zeroth}
    ds^2 = -f_0^2 dt^2 + g_0^2 dr^2 + r^2 d\Omega_{D-2}^2, \quad  f_0^2 = g_0^{-2} = 1 - \left(\frac{r_0}{r}\right)^{D-3}.
\end{equation}
We fix the value of the dilaton to be zero at infinity, and for the zeroth order solution, it is zero everywhere. The same statement holds for the axion. Therefore $\phi$ and $\chi$ are themselves at order $z$. 

The solutions (\ref{zeroth}) are parametrized by the radius of the sphere $r_0$ at the horizon.\footnote{Our $r_0$ is $\omega$ in \cite{Myers:1987qx}.} The perturbed solutions also form a one parameter family, which we will parameterize by the radius of the sphere $r_0$ at the horizon in the \emph{Einstein} frame. Note that we need to specify whether we define $r_0$ in the string frame or the Einstein frame, as the perturbed dilaton will have a nonzero value at the horizon. More explicitly, we write the spherical symmetric ansatz for the perturbed solution as
\begin{equation}\label{changeframe}
    ds^2 = e^{\frac{4}{D-2}\phi} d\tilde{s}^2 = e^{\frac{4}{D-2}\phi}\left[ - f^2 dt^2 + g^2 dr^2 + r^2 d\Omega_{D-2}^2\right] ,
\end{equation}
\begin{equation}\label{expansion}
\begin{aligned}
    f = f_0 (1+ z\mu (r)),\quad
    g  = g_0 (1+z \varepsilon(r)),\quad \phi = \phi(r), \quad \chi = \chi(r)
\end{aligned}
\end{equation}
The metric $ds^2,d\tilde{s}^2$ are the string frame and the Einstein frame metric, respectively. In (\ref{expansion}), $z$ denotes the value of $z(\phi,\chi)$ evaluated at $\phi,\chi$ being zero. Note that here we are defining the Einstein frame from the $D$ dimensional point of view, and in particular, the transformation (\ref{changeframe}) does not act on the internal space. We demand the perturbations $\mu,\varepsilon,\phi,\chi$ to be nonsingular at the horizon $r_0$ and vanish at infinity, which can be used to uniquely fix the solution as we will see.

The equations for the dilaton, axion and the metric perturbations can be decoupled by reorganizing (\ref{eom1}) - (\ref{eom3}) into
\begin{equation}\label{eomnew1}
    \nabla^2 \phi = - \frac{1}{2} z_1 Y - \frac{1}{2} z \T{^a_a} ,
\end{equation}
\begin{equation}\label{eomnew2}
     \tilde{R}_{ab} = z T_{ab} - \frac{z }{D-2} g_{ab}\left( Y + \T{^a_a}\right),
\end{equation}
\begin{equation}\label{eomnew3}
     \nabla^2 \chi = - z_2 Y,
\end{equation}
where $\tilde{R}_{ab}$ on the left hand side of (\ref{eomnew2}) is computed using the Einstein frame metric $d\tilde{s}^2$. $z_1,\, z_2$ are defined as
\begin{equation}
    z_1 \equiv z + \frac{D-2}{4}\partial_\phi z|_{\phi,\chi = 0}, \quad z_2 \equiv \partial_\chi z|_{\phi,\chi  = 0}. 
\end{equation}
To recover the tree-level case considered in \cite{Myers:1987qx}, one simply sets $z_1 = z, z_2 = 0$. 
The equation (\ref{eomnew1}) leads to
\begin{equation}\label{EOMphi}
\begin{aligned}
  &  \phi''  + \frac{(D-2) r^{D-3} - r_0^{D-3}}{r(r^{D-3} - r_0^{D-3})} \phi'  + z_1 \frac{ a_D r_0^{4(D-3)}}{ 2 r^{3D-1} (r^{D-3} - r_0^{D-3}) }\\
    & + z\frac{ ( b_D + d_D + f_D) r_0^{4(D-3)} + (c_D + e_D + g_D - b_D - d_D - f_D) r_0^{3(D-3)} r^{D-3} }{ 2 r^{3D-1} (r^{D-3} - r_0^{D-3}) } = 0.
\end{aligned}
\end{equation}
In (\ref{EOMphi}), as well as (\ref{EOMepsilon}) and (\ref{EOMmu}) below, the coefficients $a_D, b_D , ..., g_D$ are polynomials of $D$ appearing in various components of $T_{ab}$ and $Y$. We provide their explicit expressions in appendix. \ref{app:quantities}. For the metric perturbations $\varepsilon$, $\mu$, there are three equations coming from the $\tilde{R}_{tt}, \tilde{R}_{rr}$ and $\tilde{R}_{ii}$ ($i$ denotes an arbitrary angular coordinate) components of (\ref{eomnew2}), but only two of them are independent as can be shown using the Bianchi identity. We can organize them into two linear independent equations as
\begin{equation}\label{EOMepsilon}
    \varepsilon' + \frac{(D-3) r^{D-4}}{ r^{D-3}-r_0^{D-3}} \varepsilon + \frac{    (a_D + 2b_D) r_0^{4(D-3)} +(2c_D -2 b_D) r^{D-3} r_0^{3(D-3)} }{2 (D-2) r^{3D-2} (r^{D-3}-r_0^{D-3})} =0,
\end{equation}
\begin{equation}\label{EOMmu}
    -\mu' + \frac{(D-3) r^{D-4}}{ r^{D-3}-r_0^{D-3}} \varepsilon + \frac{    (a_D + 2d_D) r_0^{4(D-3)} +(2e_D -2 d_D) r^{D-3} r_0^{3(D-3)} }{2 (D-2) r^{3D-2} (r^{D-3}-r_0^{D-3})} =0.
\end{equation}

The equation (\ref{EOMphi}) for the dilaton can be integrated once to give
\begin{equation}\label{phiprime}
    \phi' = \frac{1}{r_0^7} \left[A_D \left(\frac{r_0}{r} \right)^{3D-2} + B_D \frac{r_0 (r^{2D} - r_0^{2D}) }{ r^{2D+1} \left( \frac{r^{D-3}}{r_0^{D-3}} -1 \right) }\right],
\end{equation}
where
\begin{equation}
\begin{aligned}
    A_D & = \frac{z}{2} (D-1)^2 (2D^5 - 31 D^4 + 184 D^3 -521 D^2 + 702 D- 360) + B_D, \\
    B_D & = \frac{4z - z_1}{24} (D-2) (D-3) (4D^4 - 51 D^3 + 242 D^2 - 489 D + 330). 
\end{aligned}
\end{equation}
It is possible to further integrate (\ref{phiprime}) to get a closed form expression for $\phi$, similar to what was done in the appendix. A of \cite{Callan:1988hs}. The expression is lengthy and not so illuminating, so we do not present it here. However, it is useful to note that in the tree level approximation, i.e. $z_1 = z$, the value of $\phi'$ is positive at the horizon, so the dilaton is negative near the horizon, increases and approaches zero at infinity. 
From the similarity between (\ref{eomnew1}) and (\ref{eomnew3}), we get the equation for the axion $\chi$ by replacing $z_1$ by $2z_2$ and set $z=0$ in (\ref{phiprime}). 

Next we turn to the metric perturbations. With the boundary conditions specified above, we can solve (\ref{EOMepsilon}) and (\ref{EOMmu}) and find
\begin{equation}\label{solnmu}
    \mu = - \varepsilon - \frac{C_D}{r_0^6} \left(\frac{r_0}{r} \right)^{3(D-1)} , 
\end{equation}
\begin{equation}\label{solnepsilon}
    \varepsilon = \frac{D_D}{r_0^6} \left(\frac{r_0}{r} \right)^{3(D-1)} + \frac{E_D}{r_0^6} \frac{r^{2D} - r_0^{2D} }{ r^{2D} \left( \frac{r^{D-3}}{r_0^{D-3}} -1 \right)},
\end{equation}
where\footnote{One particular difference from \cite{Myers:1987qx} is that our $E_D$ only goes as $D^5$ when $D$ is large, while the one in \cite{Myers:1987qx} goes like $D^6$. This difference will be important when we discuss the large $D$ limit of the solution in sec. \ref{sec:largeD}. }
\begin{equation}\label{cddded}
\begin{aligned}
    C_D & = \frac{2}{3} (D-1)(D-3) (2D^3 - 10D^2 + 6D + 15), \\
    D_D & = - \frac{1}{24} (D-3) (52 D^4 - 375 D^3 + 758 D^2 - 117 D - 570) , \\
    E_D & = \frac{1}{24} (D-3) (20D^4 - 225D^3 + 946 D^2 - 1779 D + 1290). 
\end{aligned}
\end{equation}

\section{Corrections to the thermodynamic quantities}\label{sec:thermodynamics}

In this section, we look at the corrections to the thermodynamic quantities of the black hole, more specifically, the temperature, mass and the entropy.  Some of these were treated in \cite{Myers:1987qx,Moura:2018psq}, while our formulas correct some exact coefficients in them. We first look at the correction to the temperature, which can be easily derived by expanding the metric around $r_0$,
\begin{equation}
\begin{aligned}
   d\tilde{s}^2 & \approx - \left. \frac{df^2}{dr} \right|_{r=r_0} (r-r_0) dt^2 + \frac{1}{ \left.\frac{d \left(\frac{1}{g^2}\right)}{dr}\right|_{r=r_0} (r-r_0) } dr^2 + r_0^2 d\Omega_{D-2}^2 \\
   & \approx - \frac{D-3}{r_0} (1+2z \left. \mu\right|_{r=r_0}) (r-r_0) dt^2 + \frac{dr^2}{  \frac{D-3}{r_0} (1-2z \left. \varepsilon\right|_{r=r_0}) (r-r_0) } + r_0^2 d\Omega_{D-2}^2 \\
   & \approx \frac{r_0}{(D-3)(1- 2 z \left. \varepsilon\right|_{r=r_0})} \left[ \frac{(D-3)^2}{4r_0^2}(1 + 2z \left.( \mu  -  \varepsilon)\right|_{r=r_0}) \rho^2 d\tau^2 + d\rho^2  \right]  + r_0^2 d\Omega_{D-2}^2.
\end{aligned}
\end{equation}
From the second line to the third line, we've done a Wick rotation and defined $\rho = 2 \sqrt{r-r_0}$. The smoothness of the solution at $r=r_0$ requires
\begin{equation}
    \tau \sim \tau + \beta, \quad \beta = \frac{4\pi r_0}{D-3} \left[1 + z \left.(\varepsilon - \mu)\right|_{r=r_0} \right].
\end{equation}
Plug in the solutions (\ref{solnmu}) and (\ref{solnepsilon}), we get the Hawking temperature
\begin{equation}\label{Hawkingtemp}
\begin{aligned}
   T & = \frac{1}{\beta} = \frac{D-3}{4\pi r_0} \left[ 1 - \left(C_D + 2  D_D + \frac{4D}{D-3} E_D \right)   \frac{z}{r_0^6}  \right]  \\
   & = \frac{D-3}{4\pi r_0} \left[ 1 - \frac{(D-1) (4D^4 - 59 D^3 + 366 D^2 - 1113 D + 1350)}{12}   \frac{z}{r_0^6}  \right].
\end{aligned}
\end{equation}

Now we turn to discuss the mass and the entropy of the black hole. These can be extracted from the free energy, which can be computed through the Euclidean action of the black hole. At the zeroth order in the $\alpha'$ expansion, the on-shell action only comes from the Gibbons-Hawking-York boundary term, which gives \cite{Gibbons:1976ue}
\begin{equation}
    I_0 = \frac{1}{D-3} \frac{\omega_{D-2} r_0^{D-2}}{4 G_N}, 
\end{equation}
where $\omega_{D-2}$ is the area of an unit $(D-2)$ sphere. To order $\mathcal{O}(z)$, the only correction to the on-shell action comes from the term $z Y(R)$ in the action,\footnote{In principle we should also consider the higher curvature correction to the boundary terms of the action. However, we expect such terms to decay as $1/r^{4D-5}$ so that they will not contribute in this calculation once we push the cutoff surface to infinity.} while the correction coming from the change of the solution vanishes due to the equations of motion. However, it is important to note that for this argument to hold, we need to have a good variational problem, namely we need to fix the size of the Euclidean circle at infinity and consider the deformation inside due to the higher curvature term. In other words, when we express the action computed this way in terms of $\beta$, we should use the relation $\beta = 4\pi r_0 / (D-3)$, but not the corrected one (\ref{Hawkingtemp}). As a conclusion, to order $z$, we have the Euclidean action
\begin{equation}
\begin{aligned}
    I & \approx I_0 - \frac{z}{16\pi G_N} \int d^D x\, \sqrt{g} \, Y(R) \\
    & = \frac{1}{D-3} \frac{\omega_{D-2} r_0^{D-2}}{4 G_N} - \frac{z \omega_{D-2} \beta }{16\pi G_N} \int_{r_0}^{\infty} dr\, r^{D-2} a_D \frac{r_0^{4(D-3)}}{r^{4(D-1)}}\\
    & = \frac{1}{D-3} \frac{\omega_{D-2} \left( \frac{(D-3)\beta}{4\pi}\right)^{D-2}}{4 G_N} \left[1 - z \frac{a_D}{3 (D-1) \left( \frac{(D-3)\beta}{4\pi}\right)^6}\right],
\end{aligned}
\end{equation}
where $a_D$ is given explicitly in appendix. \ref{app:quantities}. With the Euclidean action $I$ we can then compute the mass as
\begin{equation}
\begin{aligned}
    M & = \frac{d I}{d\beta} =   \frac{(D-2)\omega_{D-2}  \left( \frac{(D-3)\beta}{4\pi}\right)^{D-3}}{16\pi G_N} \left[1 -z \frac{ (D-8)a_D }{3 (D-1) (D-2)  \left( \frac{(D-3)\beta}{4\pi}\right)^6}\right]  \\
    & = \frac{(D-2)\omega_{D-2} r_0^{D-3}}{16\pi G_N} \left[ 1 + \frac{(D-3)(20D^4 - 225 D^3 + 946 D^2 - 1779 D + 1290)}{12}\frac{z}{r_0^6} \right]
\end{aligned}
\end{equation}
where on the second line we used (\ref{Hawkingtemp}). We can verify that we arrive at the same expression by inspecting the asymptotic form of the Einstein frame metric \cite{Callan:1988hs}, namely
\begin{equation}
\begin{aligned}
    M & = \frac{(D-2)\omega_{D-2}}{16\pi G_N} \lim_{r\rightarrow \infty} r^{D-3}\left( 1- \frac{1}{g^2}\right) =  \frac{(D-2)\omega_{D-2}}{16\pi G_N} \left[ r_0^{D-3} + 2z \lim_{r\rightarrow \infty} r^{D-3} \epsilon \right] \\
     & = \frac{(D-2)\omega_{D-2} r_0^{D-3}}{16\pi G_N} \left( 1 +2E_D \frac{z}{r_0^6} \right).
\end{aligned}
\end{equation}
Following \cite{Callan:1988hs,Myers:1987qx}, we introduce a parameter $r_{*}$ which is defined via\footnote{Our $r_*$ is the $\omega_T$ in \cite{Callan:1988hs,Myers:1987qx}}
\begin{equation}\label{rT}
   M = \frac{(D-2)\omega_{D-2} r_*^{D-3}}{16\pi G_N},\quad \textrm{or} \quad  r_{*}^{D-3} = r_0^{D-3} \left( 1 +2E_D \frac{z}{r_0^6} \right),
\end{equation}
and we can then express the Hawking temperature in terms of $r_{*}$ as
\begin{equation}\label{THmass}
     T  = \frac{D-3}{4\pi r_*} \left[ 1 - \frac{(D-8)(4D^4 - 51D^3 + 242 D^2 - 489 D + 330)}{12} \frac{z}{r_*^6}  \right]   .
\end{equation}
The benefit of this expression is that we expressed the temperature in terms of the mass, related directly via $r_{*}$. We see that the temperature correction vanishes to this order at $D=8$. When $4\leq D<8$, the temperature correction is positive, while for $D>8$, the temperature correction is negative.\footnote{One comment is that the $(D-8)$ factor in the correction term of (\ref{THmass}) can be expected beforehand. We can simply write down the ansatz $M =\frac{(D-2)\omega_{D-2} r_*^{D-3}}{16\pi G_N} $, $S = \frac{\omega_{D-2} r_*^{D-2} }{4G_N} \left[1 + \# \frac{z}{r_*^6}\right]$, and the $(D-8)$ factor will come out of $T = dM/dS$ regardless of $\#$. This suggests there is typo in (3.24) of \cite{Myers:1987qx}. }

Finally, we turn to the entropy of the black hole, which is given by
\begin{equation}
\begin{aligned}
S = \beta  \frac{dI}{d\beta} - I  & =  \frac{\omega_{D-2}  \left( \frac{(D-3)\beta}{4\pi}\right)^{D-2}}{4 G_N} \left[1 -z \frac{ (D-9)a_D }{3 (D-1) (D-3)  \left( \frac{(D-3)\beta}{4\pi}\right)^6}\right] \\
& = \frac{\omega_{D-2} r_0^{D-2} }{4G_N} \left[ 1 + (D-2) (D-3)^3 (2D-5) \frac{z}{r_0^6} \right],
\end{aligned}
\end{equation}
where we we used (\ref{Hawkingtemp}) to get to the second line. It is interesting to also express the entropy in terms of the more physical parameter $r_*$ via (\ref{rT}), and we find
\begin{equation}\label{Entropymass}
    S = \frac{\omega_{D-2} r_*^{D-2} }{4G_N} \left[ 1 + \frac{(D-2) (4D^4 - 51 D^3 + 242 D^2 - 489 D + 330)}{12} \frac{z}{r_{*}^6}\right].
\end{equation}

We arrive at the conclusion that for $D\geq 4$, the stringy correction gives a positive contribution to the entropy of a black hole with a fixed mass. This is the same qualitative feature found in the bosonic string theory \cite{Callan:1988hs}.

\section{Comments on the large $D$ limit}\label{sec:largeD}

Based on earlier observations in \cite{Emparan:2013xia}, it was argued in \cite{Chen:2021emg} that in the large $D$ limit, the stringy corrections to the near horizon geometry of the Schwarzschild black hole can be captured by the two dimensional $SL(2,R)_k/U(1)$
WZW model. A preliminary check of the type II superstring case was presented in the paper. Here we expand on the check using the solutions we found.
The large $D$ limit involves taking $D$ to infinity while keeping $r_0/D$ finite. By the argument in \cite{Chen:2021emg}, the ratio $r_0/D$ in the large $D$ limit should be related to the level $k$ of the supersymmetric gauged WZW model (see \cite{Giveon:2003wn} and references therein) through 
\begin{equation}\label{levelr0D}
     k = \left( \frac{2r_0}{D}\right)^2,
\end{equation}
which should hold to all orders in the $\alpha'$ expansion. We set $\alpha'=1$ in this section. In the supersymmetric gauged WZW model, the temperature of the black hole is related to $k$ via
\begin{equation}\label{TH2d}
    T = \frac{1}{2\pi \sqrt{k}}. 
\end{equation}
Combining (\ref{levelr0D}) and (\ref{TH2d}), we conclude that in the large $D$ limit 
\begin{equation}
    T = \frac{D}{4\pi r_0},
\end{equation}
namely all the $\alpha'$ corrections should vanish. We see that this is consistent with our formula (\ref{Hawkingtemp}), where the correction goes like $D^5/r_0^6$, which vanishes when we take the large $D$ limit. A subtlety in this argument is that in the argument of \cite{Chen:2021emg}, $r_0$ was defined in the string frame, while in (\ref{Hawkingtemp}), $r_0$ was defined in the Einstein frame. This is not an issue here since the two definitions differ by a factor of $\exp\left( \frac{2\phi}{D-2}\right)$, which becomes one in the large $D$ limit ($\phi$ is order one as we will discuss below). In other words, the frames don't matter as long as we are not dealing with powers of $r_0$ that scale as $D$. 

Unlike the bosonic case \cite{Dijkgraaf:1991ba,Tseytlin:1991ht,Tseytlin:1993df}, it was shown in  \cite{Jack:1992mk,Tseytlin:1993my} that the metric and the dilaton of the supersymmetric two dimensional black hole are uncorrected under the stringy corrections, and take the form
\begin{equation}\label{metric2d}
    ds^2 = k \left( d\rho^2 + \tanh^2 \rho d\theta^2\right), \quad e^{-2\phi_{2d}}  = e^{-2\phi_0} \cosh^2 \rho. 
\end{equation}
To connect it with the large $D$ metric, we note that the two dimensional dilaton can be interpreted as coming from the volume of the $(D-2)$ sphere in the Einstein frame, which goes like $r^{D-2}$. In other words, in the large $D$ limit, we identify
\begin{equation}
    \left( \frac{r}{r_0}\right)^{D} = \cosh^2 \rho.
\end{equation}
With this identification, (\ref{metric2d}) matches the zeroth order metric (\ref{zeroth}) in the large $D$ limit  given (\ref{levelr0D}).  From (\ref{solnmu}) and (\ref{solnepsilon}), we see that the corrections to the metric in the Einstein frame vanish in the large $D$ limit, namely
\begin{equation}
    \mu ,\,\, \varepsilon \sim \mathcal{O}\left( \frac{1}{D}\right).
\end{equation}
In other words, in the large $D$ limit there is no $\mathcal{O}(\alpha'^3)$ corrections to the Einstein frame metric. The string frame metric differs from the Einstein frame by a factor $\exp\left( \frac{4\phi}{D-2}\right)$ which becomes one in the large $D$ limit. As a conclusion, if we consider the string frame metrics on both sides, the two dimensional part of the large $D$ metric matches with the 2d theory to order $\mathcal{O}(\alpha'^3)$.

The dilaton correction is nonvanishing in the large $D$ limit. In fact, by integrating (\ref{phiprime}), it goes like
\begin{equation}
   \phi =  - \frac{z }{3} \left( \frac{D}{r_0}\right)^6 \left( \frac{r_0}{r}\right)^{3D}, \quad D\rightarrow\infty. 
\end{equation}
One consequence of a nonzero dilaton is that if we instead choose to define the radial coordinate in the string frame, denoted by $r_s$, its relation with the coordinate $\rho$ \emph{will} be modified by the stringy correction through the relation
\begin{equation}
  e^{-2(\phi  - \left.\phi\right|_{r_{0,s}})}  \left(\frac{r_s}{r_{0,s}}\right)^D = \cosh^2 \rho. 
\end{equation}

As a summary, we've verified that the $\mathcal{O}(\alpha'^3)$ stringy corrections to the black hole in the large $D$ limit is consistent with that in the two dimensional supersymmetric gauged WZW model.

\paragraph{Acknowledgement}
We would like to thank Juan Maldacena for helpful discussions and comments on the manuscript. The general relativity computations in this paper were done with the help of the mathematica package from Tom Hartman's webpage. 

\appendix

\section{List of some relevant quantities}\label{app:quantities}

Here we provide the analytic expressions of some geometric quantities. All the following quantities are evaluated on the zeroth order solution (\ref{zeroth}). 

\begin{equation}
  \R{^t^r_t_r} = \frac{(D-3)(D-2)}{2} \frac{r_0^{D-3}}{r^{D-1}},\quad  \R{^t^i_t_j} =\R{^r^i_r_j}  = - \frac{D-3}{2}  \frac{r_0^{D-3}}{r^{D-1}}\delta^{i}_{j},\quad  \R{^i^j_k_l} =   \frac{r_0^{D-3}}{r^{D-1}} (\delta^{i}_{k} \delta^{j}_{l} - \delta^{i}_{l} \delta^j_k).
\end{equation}
\begin{equation}
 Y(R) = a_D \frac{r_0^{4(D-3)}}{r^{4(D-1)}},
\end{equation}
\begin{equation}
    \tensor{T}{^t_t} = \frac{r_0^{3(D-3)} \left[ b_D (r_0^{D-3} - r^{D-3}) + c_D r^{D-3} \right]}{r^{4(D-1)}},
\end{equation}
\begin{equation}
    \tensor{T}{^r_r} = \frac{r_0^{3(D-3)} \left[ d_D (r_0^{D-3} - r^{D-3}) + e_D r^{D-3} \right]}{r^{4(D-1)}},
\end{equation}
\begin{equation}\label{Tii}
      \tensor{T}{^i_i} = \frac{r_0^{3(D-3)} \left[ f_D (r_0^{D-3} - r^{D-3}) + g_D r^{D-3} \right]}{r^{4(D-1)}}.
\end{equation}
In the above expressions, indices $i,j,k,l$ denote the angular directions. The index $i$ is summed over in (\ref{Tii}). The polynomials $a_D,b_D, ...,g_D$ are computed by evaluating the quantities explicitly with mathematica for dimensions $D= 4, 5, ..., 12$, and then use polynomials to interpolate the coefficients. The answers are uniquely determined since they can at most be polynomials with degree 8. They are given by
\begin{equation}
\begin{aligned}
a_D & = \frac{1}{4} (D-1)(D-2)(D-3) (4D^4 - 51 D^3 + 242 D^2 - 489 D +330) \\
b_D & = \frac{3}{2} (D-1) (D-2) (D-3)^2 (4D^3 - 15D^2 -2D +25) \\
 c_D & = \frac{1}{2}(D-1)(D-2)(D-3)^3 (7D-25) \\
  d_D & = \frac{1}{2} (D-1) (D-2) (D-3) (4D^4 - 33D^3 + 65 D^2 + 57 D - 165)\\
  e_D & = c_D \\
   f_D & =  -3 (D-1 ) (D-2) (D-3) (2D^5 - 21D^4 + 75 D^3 - 102 D^2 + 49 D - 15) \\
    g_D & = - (D-1) (D-2) (D-3) (2D^5 - 25D^4 + 113 D^3 -  216 D^2 + 165 D - 75). 
\end{aligned}
\end{equation}

\section{Derivation of $\frac{\delta Y}{\delta g^{ab}}$}\label{app:variation}

Here we present some details of the computation of $\frac{\delta Y}{\delta g^{ab}}$. We have not tried to optimize the following derivation. By using the Bianchi identity $\R{^a_{(b}_c_{d)}}=0$ and exchanging labels, we can write $Y$ as
\begin{equation}
\begin{aligned}
    Y & =  2 R_{hmnk} \R{_p^m^n_q } R^{hwsp}\R{^q_w_s^k} + R_{hkmn} \R{_p_q^m^n} R^{hwsp}\R{^q_w_s^k}\\
&   = 4 \R{^h_w_s_p}\R{_q^w^s^k}R_{hmnk} \R{^p^m^n^q } - 2  \R{^h_w_s_p}\R{_q^w^s^k}R_{hmnk} \R{^p^n^m^q } \\
& = 4 Y_1 - 2Y_2,
\end{aligned}
\end{equation}
where
\begin{equation}
    Y_1 \equiv  \R{^h_w_s_p}\R{_q^w^s^k}R_{hmnk} \R{^p^m^n^q } , \quad Y_2 \equiv \R{^h_w_s_p}\R{_q^w^s^k}R_{hmnk} \R{^p^n^m^q }. 
\end{equation}
We have
\begin{equation}
\begin{aligned}
    \delta Y_1 & = \delta\left(   \R{^h_w_s_p}\R{_q^w^s^k}R_{hmnk} \R{^p^m^n^q } \right)\\
    & = \delta\R{^h_w_s_p}\R{_q^w^s^k}R_{hmnk} \R{^p^m^n^q }  + \R{^h^w^s_p}\delta\R{_q_w_s^k}R_{hmnk} \R{^p^m^n^q } \\
    & \quad + \R{_h_w_s_p}\R{_q^w^s^k} \delta \R{^h_m_n_k} \R{^p^m^n^q } + \R{^h_w_s^p}\R{_q^w^s^k} \R{_h^m^n_k} \delta\R{_p_m_n^q }\\
    & \quad +  \delta g^{aw} \R{^h_w_s_p} \R{_q_a^s^k}R_{hmnk} \R{^p^m^n^q } + \delta g^{as} \R{^h_w_s_p} \R{_q^w_a^k}R_{hmnk} \R{^p^m^n^q } \\
    & \quad + \delta g_{ah} \R{^h_w_s_p}\R{_q^w^s^k} \R{^a_m_n_k} \R{^p^m^n^q } + \delta g^{ap} \R{^h_w_s_p}\R{_q^w^s^k}R_{hmnk} \R{_a^m^n^q } \\
    & \quad + \delta g^{am} \R{^h_w_s_p}\R{_q^w^s^k}R_{hmnk} \R{^p_a^n^q } + \delta g^{an} \R{^h_w_s_p}\R{_q^w^s^k}R_{hmnk} \R{^p^m_a^q } 
\end{aligned}
\end{equation}
The reason that we have several $\delta g$ terms is because we lowered or raised some indices of the Riemann tensor under the variation, so that we can bring it to the following simpler form:
\begin{equation}\label{deltaY1}
\begin{aligned}
    \delta Y_1 & = \delta \R{^h_w_s_p}\left[ 2\R{_q^s^w^k}\R{_h_m_n_k} \R{^p^n^m^q}  + 2\R{_q^w^s^k} \R{_h_m_n_k}  \R{^p^m^n^q}  \right]\\
    & \quad + 2\delta g^{ab} \left(  \R{^h_b_s_p}\R{_q_a^s^k}R_{hmnk} \R{^p^m^n^q } +  \R{^h_b_s_p}\R{^k_a^s_q}R_{hmnk} \R{^q^m^n^p } \right).
\end{aligned}
\end{equation}
We can do a similar manipulation for $Y_2$,
\begin{equation}
\begin{aligned}
    \delta Y_2 & = \delta \left( \R{^h_w_s_p}\R{_q^w^s^k}\R{_h_m_n_k} \R{^p^n^m^q }  \right) \\
    & =  \delta\R{^h_w_s_p}\R{_q^w^s^k}R_{hmnk} \R{^p^n^m^q } + \R{^h^w^s_p} \delta\R{^q_w_s_k}\R{_h_m_n^k} \R{^p^n^m^q } \\
    & \quad + \R{_h_w_s_p}\R{_q^w^s^k} \delta \R{^h_m_n_k} \R{^p^n^m^q } + \R{^h_w_s^p}\R{_q^w^s^k}\R{_h^m^n_k} \delta\R{_p_n_m^q } \\
    & \quad + \delta g_{aq} \R{^h_w_s_p}\R{^a^w^s^k}\R{_h_m_n_k} \R{^p^n^m^q } + \delta g^{aw} \R{^h_w_s_p}\R{_q_a^s^k}\R{_h_m_n_k} \R{^p^n^m^q } \\
    & \quad + \delta g^{as} \R{^h_w_s_p}\R{_q^w_a^k}\R{_h_m_n_k} \R{^p^n^m^q }  + \delta g^{ak} \R{^h_w_s_p}\R{_q^w^s_a}\R{_h_m_n_k} \R{^p^n^m^q } \\
    & \quad + \delta g_{ah} \R{^h_w_s_p}\R{_q^w^s^k}\R{^a_m_n_k} \R{^p^n^m^q } + \delta g^{ap} \R{^h_w_s_p}\R{_q^w^s^k}\R{_h_m_n_k} \R{_a^n^m^q }  \\
    & \quad + \delta g^{an} \R{^h_w_s_p}\R{_q^w^s^k}\R{_h_m_n_k} \R{^p_a^m^q } + \delta g^{am}\R{^h_w_s_p}\R{_q^w^s^k}\R{_h_m_n_k} \R{^p^n_a^q }
\end{aligned}
\end{equation}
and get
\begin{equation}\label{deltaY2}
    \delta Y_2 = 4 \delta\R{^h_w_s_p}\R{_q^w^s^k}R_{hmnk} \R{^p^n^m^q } + 4\delta g^{ab} \R{^h_b_s_p}\R{_q_a^s^k}R_{hmnk} \R{^p^n^m^q }
\end{equation}
Combining (\ref{deltaY1}) and (\ref{deltaY2}), we get
\begin{equation}
\begin{aligned}
    \delta Y & = 8\delta \R{^h_w_s_p}  \left[ \R{_q^s^w^k}\R{_h_m_n_k} \R{^p^n^m^q}  + \R{_q^w^s^k} \R{_h_m_n_k}  \R{^p^m^n^q}  - \R{_q^w^s^k}R_{hmnk} \R{^p^n^m^q }  \right]\\
    & \quad + 8\delta g^{ab} \left(  \R{^h_b_s_p}\R{_q_a^s^k}R_{hmnk} \R{^p^m^n^q } +  \R{^h_b_s_p}\R{^k_a^s_q}R_{hmnk} \R{^q^m^n^p } - \R{^h_b_s_p}\R{_q_a^s^k}R_{hmnk} \R{^p^n^m^q }\right) \\
    & = 8\delta \R{^h_w_s_p} \F{_h^w^s^p} +  8\delta g^{ab} H_{ab},
\end{aligned}
\end{equation}
where $F$ and $H$ are the tensors in the square bracket and the round bracket, respectively.
Now we use 
\begin{equation}
    \delta \R{^\rho_\mu_\lambda_\nu} = \nabla_\lambda \left( \delta \tGamma{^\rho_\nu_\mu}\right) - \nabla_\nu \left( \delta \tGamma{^\rho_\lambda_\mu}\right), \quad \delta \tGamma{^\rho_\mu_\nu} = \frac{1}{2}  g^{\rho\sigma} \left( \nabla_\nu \delta g_{\mu\sigma} + \nabla_\mu \delta g_{\nu\sigma} - \nabla_{\sigma} \delta g_{\mu\nu}\right)
\end{equation}
and integrate by part to get
\begin{equation}
\begin{aligned}
    \delta Y  = 8\delta g^{ab} & \left\{ H_{ab} + \frac{1}{2} \left[\nabla_s, \nabla_p\right] \F{_a_b^s^p} - \frac{1}{2} \nabla_w \nabla_s \F{_a^w^s_b} + \frac{1}{2} \nabla_h\nabla_s \F{^h_a^s_b}  \right. \\
    & \quad \left. + \frac{1}{2} \nabla_w \nabla_p \F{_a^w_b^p} - \frac{1}{2} \nabla_h \nabla_p \F{^h_a_b^p} \right\} + \textrm{total derivatives}.
\end{aligned}
\end{equation}
From here it is straightforward to simplify the expression into (\ref{variationformula}). To get (\ref{variationformula}), it is useful to note that we can symmetrize the expression in the bracket under $a\leftrightarrow b$.

\bibliographystyle{JHEP}
\bibliography{cite}

\end{document}